# A strain tunable single-layer MoS$_2$ photodetector

*Patricia Gant[1,*], Peng Huang[1,2], David Pérez de Lara[3], Dan Guo[2], Riccardo Frisenda[1,*], Andres Castellanos-Gomez[1,*]*

[1]Materials Science Factory, Instituto de Ciencia de Materiales de Madrid (ICMM-CSIC), Madrid E-28049, Spain.
[2]State Key Laboratory of Tribology, Tsinghua University, Beijing 100084, China.
[3]Instituto Madrileño de Estudios Avanzados en Nanociencia (IMDEA-nanociencia), Campus de Cantoblanco, Madrid, Spain.

*E-mail: patricia.gant@csic.es, riccardo.frisenda@csic.es , andres.castellanos@csic.es

**ABSTRACT:** Strain engineering, which aims to tune the bandgap of a semiconductor by the application of strain, has emerged as an interesting way to control the electrical and optical properties of two-dimensional (2D) materials. Apart from the changes in the intrinsic properties of 2D materials, the application of strain can be also used to modify the characteristics of devices based on them. In this work, we study flexible and transparent photodetectors based on single-layer MoS$_2$ under the application of biaxial strain. We find that by controlling the level of strain, we can tune the photoresponsivity (by 2-3 orders of magnitude), the response time (from <80 ms to 1.5 s) and the spectral bandwidth (with a gauge factor of 135 meV/% or 58 nm/%) of the device.

**Introduction:**

Tuning the characteristics of circuit components with an external knob is at the deep core of modern electronics. Good examples of that are the field-effect transistors whose conductance can be adjusted by means of an applied voltage to the gate electrode.[1–5] The development of new tuning knobs have undoubtedly opened up possibilities to design new device concepts.[6–11] Strain engineering provides a powerful route to modify the electrical and optical properties of electronic materials and thus it has the potential to become an excellent external tuning knob.[12–15] Conventional strain engineering approaches, however, typically yield fixed strain levels. This has radically changed with the isolation of 2D materials which provide an excellent platform for strain engineering as they can be easily stretched and bended to a large extent in a reversible way.[16–20] Moreover, optical spectroscopy techniques have demonstrated that tensile uniaxial or biaxial strain effectively decreases the bandgap in atomically thin transition metal dichalcogenides.[21–27] Previous works on 2D-based flexible photodetectors subjected to mechanical deformations don't provide any direct measurement that ensures that the deformation of the substrate is effectively translated into strain in the 2D material and they don't characterize the effect of the mechanical deformation of the photodetector on their photocurrent spectra.[28,29] Drawing conclusions about the strain dependent performance of 2D based photodetectors from those results is thus challenging.

Here we exploit the strain tunable bandgap of single-layer molybdenum disulfide ($MoS_2$) to fabricate a photodetector device whose responsivity, response time and spectral bandwidth can be adjusted by an externally applied biaxial tensile or compressive strain. We find a remarkably large strain sensitivity of the cut-off wavelength of 58 nm/% of strain (~135 meV/%) making it possible to extend the detection spectral bandwidth with respect to pristine unstrained devices. The method presented here are general and can be applied to photodetectors based on other 2D materials. The case of black phosphorus, for example, could be particularly interesting as a larger strain dependent band gap tuning and an opposite strain dependence is expected. [30,31]

**Results and discussion:**

We fabricate single-layer $MoS_2$ photodetectors with a simple photocell (or photoresistor) geometry (Figure 1a). Single-layer $MoS_2$ is prepared by mechanical exfoliation of bulk molybdenite (see Materials and Methods for details) and the exfoliated flakes are then transferred to a Gel-Film (from Gel-Pak®) substrate. The single-layer regions are identified by micro-reflectance spectroscopy [32,33] and quantitative analysis of transmission mode optical microscopy images. The selected single-layer flakes are then transferred onto a polycarbonate (PC) substrate with pre-patterned drain-source electrodes by a dry-transfer deterministic placement method [34,35] and are annealed at 100ºC to improve the electrical contact between the flake and the electrodes. Figure 1b shows an optical microscopy image of one of the assembled single-layer $MoS_2$ photodetector devices fabricated onto PC. Note that it has been previously demonstrated that polymeric substrates, such as PC, polyimide and polydimethylsiloxane (PDMS) can be used to integrate photonic devices. [36–38] Polycarbonate has been used as substrate because of the trade-off between high Young's modulus and large thermal expansion that allows one to biaxially stretch (or compress) the $MoS_2$ by moderately warming up (or cooling down) the substrate while ensuring an excellent and homogeneous strain transfer [39] (see Supplementary Material Figure S1 to observe the spatial homogeneity of the transferred strain). Substrates with larger thermal expansion but lower Young's modulus cannot effectively transduce their thermal expansion into biaxial strain, as predicted by finite elements analysis (see Supplementary Material Figure S2). In the case of polycarbonate ($E$ = 2.5 GPa) the calculated efficiency is larger than 80%. The mechanical model, which lacks atomistic details, gives only a coarse estimation of the strain transfer efficiency and for this reason we assume in the rest of the paper a perfect transduction of thermal expansion to biaxial strain. Note that by assuming perfect transduction, the gauge factors obtained in this work can be considered as lower bound limits. Figure 1c shows differential reflectance spectra acquired on a single-layer $MoS_2$ device at different tensile strain levels from 0% (substrate at room temperature, $T$ = 25ºC) to 0.48% (substrate heated at $T$ = 100ºC). The A and B exciton peaks in the reflectance spectra redshift upon increasing the substrate temperature (and thus the biaxial tensile strain) indicating a narrowing of the $MoS_2$ bandgap. Figure 1d shows the relationship between the substrate temperature increase and its biaxial expansion. We address the reader to Reference [39] and the Supplementary Material (Figure S3) for details about the measurement of thermal expansion and the calibration of the applied biaxial strain.

In order to test the spectral response of the $MoS_2$ photocell detectors we illuminate the devices with a tunable light source (Bentham TLS120Xe) to select the wavelength (with a full-width at half maximum of ~10 nm) while the current across the device (biased at 10 V) is measured. The light is focused into a spot of 400 µm of diameter with a power density of 8 mW/cm$^2$. At each step during the wavelength sweep we measure both the dark and illuminated current to rule out drifts during the measurement. The responsivity *R* of the device can be extracted from the photocurrent values as

$$R = \frac{I_{ph}}{P_{dens} \cdot A_{dev}}$$

where $P_{dens}$ is the incident light power density and $A_{dev}$ is the area of the MoS$_2$ channel. The biaxial strain applied to the device is controlled through a thermal stage (see Materials and Methods).

The responsivity spectrum measured for the unstrained device is comparable with the data reported in literature for MoS$_2$ phototransistors fabricated by electron beam lithography on SiO$_2$/Si substrates at similar illumination power density and biasing conditions (larger responsivities are reported for devices at very low illumination levels and upon much larger drain-source electric field biasing conditions).[40–42] The spectra clearly show two peak features, which are in good agreement with the B and A excitonic resonances also observed in the reflection spectra of single-layer MoS$_2$,[43–45] and an abrupt drop of the device responsivity after the A exciton peak. When the polycarbonate substrate is biaxially strained, the MoS$_2$ photocell responsivity spectrum changes significantly. In Figure 2a, the overall responsivity values increase by a factor of ~100 when the strain is increased from -0.8% to 0.48% (a factor of ~1000 for device #2 where the strain ranges from -1.44% to 0.48%). We attribute this effect to the applied strain as similar photocells fabricated on SiO$_2$ (which have negligible thermal expansion) do not show this strong enhancement of the photoresponse (see a comparison between devices fabricated on SiO$_2$/Si and on PC in the Supplementary Material, Figures S4 and S5).

To get an insight about the tensile strain induced increase of the overall responsivity we studied the response time of the devices. For $MoS_2$ photodetectors there are two dominant mechanisms for the photocurrent generation (photoconductance and photogating) that can be easily identified through the response time of the devices.[40,46–48] In the photoconductance, the photoexcited electrons and holes are separated through the bias voltage leading to an increase of the current flowing through the semiconductor channel. The typical response time of photoconductive devices is <10 ms. In the photogating mechanism the photoexcited electrons are drifted by the bias voltage to the drain electrode (with a typical timescale of ~10-100 ns) and holes are trapped (with a typical timescale of 10 ms – 10 s). Due to charge conservation in the channel one new electron has to jump from the source into the channel once the drifting electron reaches the drain electrode, leading to a photoconductive gain that is proportional to the ratio between the electron drifting time and the hole trapping time (which can be extracted from the response time of the device to pulsed illumination). Therefore, photoconductive-dominated devices have a fast response but low responsivity values while photogating-dominated ones are typically slow but present ultrahigh values of responsivity.

The response time to pulsed illumination for devices under compressive strain is very sharp (the response is faster than that of our experimental setup, 80 ms, compatible to what one would expect for a photoconductive-dominated device). Tensile strained devices, on the other hand, show much slower response times ~1.5 s (Figure 2a inset) indicating the presence of long-lived charge traps that could explain the large increase of responsivity upon tension through the gain associated to the photogating mechanism. We rule out the effect of the temperature as photodetectors fabricated on $SiO_2$, with negligible thermal expansion, do not show a sizeable variation of the response time in the temperature range of 25ºC to 100ºC (see Figure S6 in the Supplementary Material). Therefore, biaxial strain seems to be responsible of a change of the photocurrent generation mechanism from photoconductive (for compressive strain) to photogating (for tensile strain), although the microscopic mechanism is still unknown, and it will be subject of further study. Note that we also studied the effect of strain on the Schottky barrier height through scanning photocurrent (see Figures S7 and S8 in the Supplementary Material) finding that the Schottky barrier height in our devices is very small (~14 meV for pristine unstrained devices) and they show a moderate decrease of the barrier height upon tensile straining. This mechanism, although small, also contributes to the increase of photoresponse observed for tensile stressed devices.

The excitonic features, visible in all the photoresponse spectra of strained $MoS_2$, are redshifted (blueshifted) when increasing the tensile (compressive) strain value with a strain gauge factor of 31 nm/% (~94 meV/%) for exciton A (inset Figure 2b). The shift of the exciton peaks with strain in the responsivity spectra is in good agreement with predictions based on density functional theory calculations,[49] with the shift observed in differential reflectance (Figure 1c and Ref. [39]) and photoluminescence measurements on pressurized blisters of $MoS_2$ [27]. The observed shift of the A exciton upon straining (for 3 different devices) with a constant gauge factor over the whole strain range studied here indicates that strain is being transferred to the single-layer $MoS_2$ flake without slippage. Moreover, one might wonder about the presence of buckling upon compressive strain but the critical (or minimum) strain that is necessary for the buckling to occur is $\varepsilon_c$

= $0.25 \cdot (3 \cdot E_{\text{substrate}}/E_{\text{flake}})^{2/3}$.[50,51] In our case a compressive strain of ~ -2.7% is needed for buckling to occur.

The cut-off wavelength is shifted as well upon straining, as can be seen in Figure 2b. The device shows a strain gauge factor of the cut-off wavelength in the range of ~58 nm/% (~135 meV/%) of biaxial strain. Therefore, we demonstrated that applying tensile biaxial strain to the $MoS_2$ device can be an effective strategy to increase both the responsivity and the wavelength bandwidth of the photodetector (at the expense of a slower response time) while compressive strain can be exploited to yield faster photodetectors (although with a lower photoresponse and with a narrower wavelength bandwidth). This adaptable optoelectronic performance of this device can be very useful to adjust the photodetector operation to different lighting conditions, similarly to human eye adaptability (scotopic vision during the night vs. photopic vision during the daylight).[52]

A direct consequence of the strain induced redshift for tensile biaxial strain is that one can achieve a sizeable response for wavelengths even beyond the cut-off of pristine (unstrained) $MoS_2$. Figure 2c shows an example where the photocurrent of the device (with light of 740 nm with a power density of 5 mW/cm$^2$, applying a bias voltage of 5 V) is measured while changing the strain level in time. One can appreciate how the photocurrent at 740 nm increases substantially upon increasing the strain level of the device (from 0.16% to 0.48% of tensile biaxial strain) in a reproducible way, which can be attributed to the strain induced redshift of the device cut-off wavelength. Interestingly, the device can be strain tuned rather quickly (in ~20 seconds time scale, see Figure 2d) and this tuning time most likely is only currently limited by the thermalization time of our temperature stage and it could be further improved by employing micro-heaters located underneath the photocell device.

In order to study the reproducibility of the redshift in the responsivity spectra of single-layer $MoS_2$ photodetectors, several cycles of application/release of tensile strain were performed in another $MoS_2$ photodetector. Figure 3 shows the device responsivity for 740 nm illumination wavelength (power density of 12 mW/cm$^2$ and bias voltage of 10 V) recorded during some of the straining cycles. We observe how the responsivity evolves from negligible values (within the experimental setup noise level) for the unstrained device towards increasingly high values for the tensile strained devices, similarly to what was displayed for few cycles in Figure 2c. The device shows a good reproducibility during the whole range of cycles (up to 40 cycles were applied).

**Conclusions:**

In summary we have harnessed the strain tunability of the band structure of single-layer $MoS_2$ to fabricate photodetectors with strain actuated bandwidth. The strain in these devices can be reversibly applied through the thermal expansion (shrinkage) of their substrate material which induces tensile (compressive) biaxial stress. We find that upon tensile straining the photoresponse increases and that the excitonic features present in the spectra redshift increasing the bandwidth in agreement with previous spectroscopic works. Interestingly the spectra also develop a slowly decaying tail for long wavelengths which further increases the bandwidth. We extract a strain gauge factor for the wavelength cut-off shift of up to ~58 nm/% (~135 meV/%). This remarkably large value demonstrates that 2D semiconductors hold a great promise for future straintronic devices where strain is employed as a variable tuning knob. Indeed, the possibility of strain tuning the optoelectronic performance of photodetector devices in a fast timescale opens up the possibility to fabricate artificial photonic devices that mimic the adaptability of the human eye. The photodetectors discussed in this work, which can be tuned from a fast, low responsivity and narrowband state to a slow, very sensitive and wideband one, can open the possibility of developing adaptable artificial photonic elements. Indeed, the method presented here can be easily translated to fabricate strain tunable photodetectors based on other 2D materials.

**Materials and Methods:**

*Materials*

The $MoS_2$ flakes have been obtained by mechanical exfoliation with Nitto tape (SPV 224) from a bulk natural crystal (Moly Hill mine, Quebec, QC, Canada) onto a viscoelastic polydimethylsiloxane (PDMS) stamp from Gel-Pak® (Gel-Film WF x4 6.0mil).

*Device fabrication*

The $MoS_2$ flakes located on the PDMS were characterized by optical transmission microscopy (Motic BA310Met-T metallurgical microscope, equipped with a 18 megapixel digital camera AMScope MU1803 and a fiber-coupled compact spectrometer Thorlabs CCS200/M) to determine the number of layers.[32] The single-layer $MoS_2$ flakes were then transferred between pre-patterned Au/Ti electrodes (fabricated by electron-beam evaporation through a metal shadow mask from Ossila®, part number

E324) on polycarbonate (PC) substrate by using an all dry deterministic transfer method.[34]

*Strain application*

The biaxial strain on the photodetector devices is achieved through the thermal expansion of the PC substrate, controlled by a thermal stage (Linkam HFS600-P for the measurements shown in Figure 2 and a Peltier element with an approximated consumption of 5 W to reach the highest temperature value for the rest of the measurements).

*Photocurrent spectroscopy measurements*

Photodetectors #1, #2 and #3 were measured in a pure $N_2$ atmosphere, the rest of the devices were measured under ambient conditions (average humidity 20%). The current *vs.* voltage characteristics of the devices were measured with a Keithley 2450 source-meter unit while the devices were illuminated focusing the light coming from a fiber coupled light source into a 400 μm spot, covering the whole area of the device and providing a homogeneous power density. The cut-off wavelength is extracted from the wavelength value at which the photocurrent drops below the setup noise level (1pA). A Bentham TLS120Xe tunable light source was used for the measurements shown in Figure 2, a halogen lamp equipped with a VariSpec™ Liquid Crystal Tunable Filter was used for the measurement shown in Figure S6 in the Supporting Information. The rest of the photocurrent measurements were carried out with high power fiber-coupled LED sources from Thorlabs. Figure S9 in the Supplementary Material shows an example of current vs.

voltage characteristics and of response time measurements acquired at different strain levels.

**Supplementary Material**

Scanning micro-reflectance maps; finite element analysis to determine the strain transfer; strain calibration; disentangling temperature change from biaxial strain; scanning photocurrent measurements; examples of strain tuning of the current vs. voltage characteristics and response time measurements; characteristics of other devices (Figures S10 and S11); strain tuning of the power-dependent photocurrent generation (Figure S12).

**Acknowledgements**

This project has received funding from the European Research Council (ERC) under the European Union's Horizon 2020 research and innovation programme (grant agreement n° 755655, ERC-StG 2017 project 2D-TOPSENSE). ACG and PG acknowledge funding from the EU Graphene Flagship funding (Grant Graphene Core 2, 785219). RF acknowledges support from the Netherlands Organization for Scientific Research (NWO) through the research program Rubicon with project number 680-50-1515. DPdL acknowledges support from MINECO through the program FIS2015-67367-C2-1-p.

**Data availability**

The raw/processed data required to reproduce these findings cannot be shared at this time due to technical or time limitations.

Data will be made available on request.

# Figures

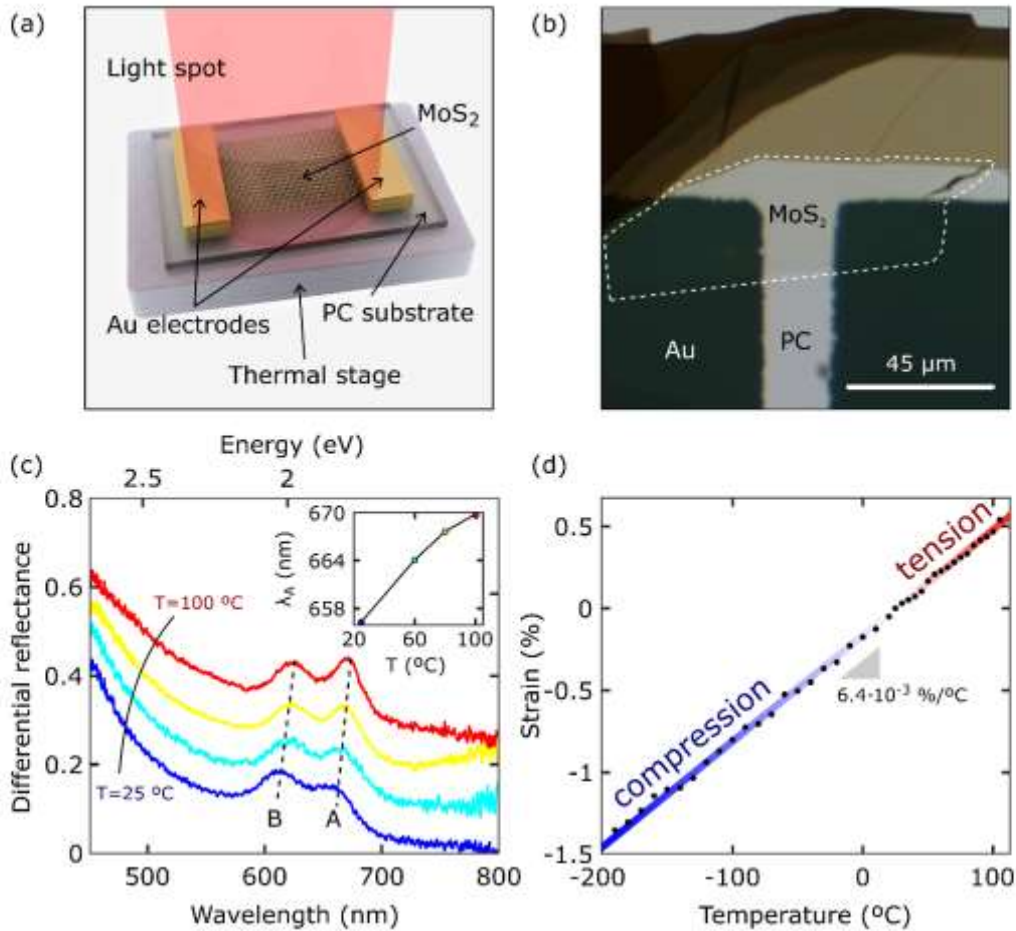

**Figure 1.** (a) Schematic picture of the setup used to perform all the measurements. The MoS$_2$ photodetectors are placed on a thermal stage and illuminated from the top. (b) Optical transmission photograph of a single-layer MoS$_2$ photodetectors fabricated on polycarbonate (PC). (c) Differential reflectance spectra measured at different temperatures (vertically shifted by 0.08 to facilitate their comparison). Inset: wavelength of the exciton A as a function of temperature. (d) Calibration of the polycarbonate expansion/compression as a function of the temperature in the range from -200ºC (corresponding to a compression of -1.48%) to +100ºC (corresponding to a tension of + 0.48%). The line in (d) indicates the best-fit of the experimental data to a linear trend.

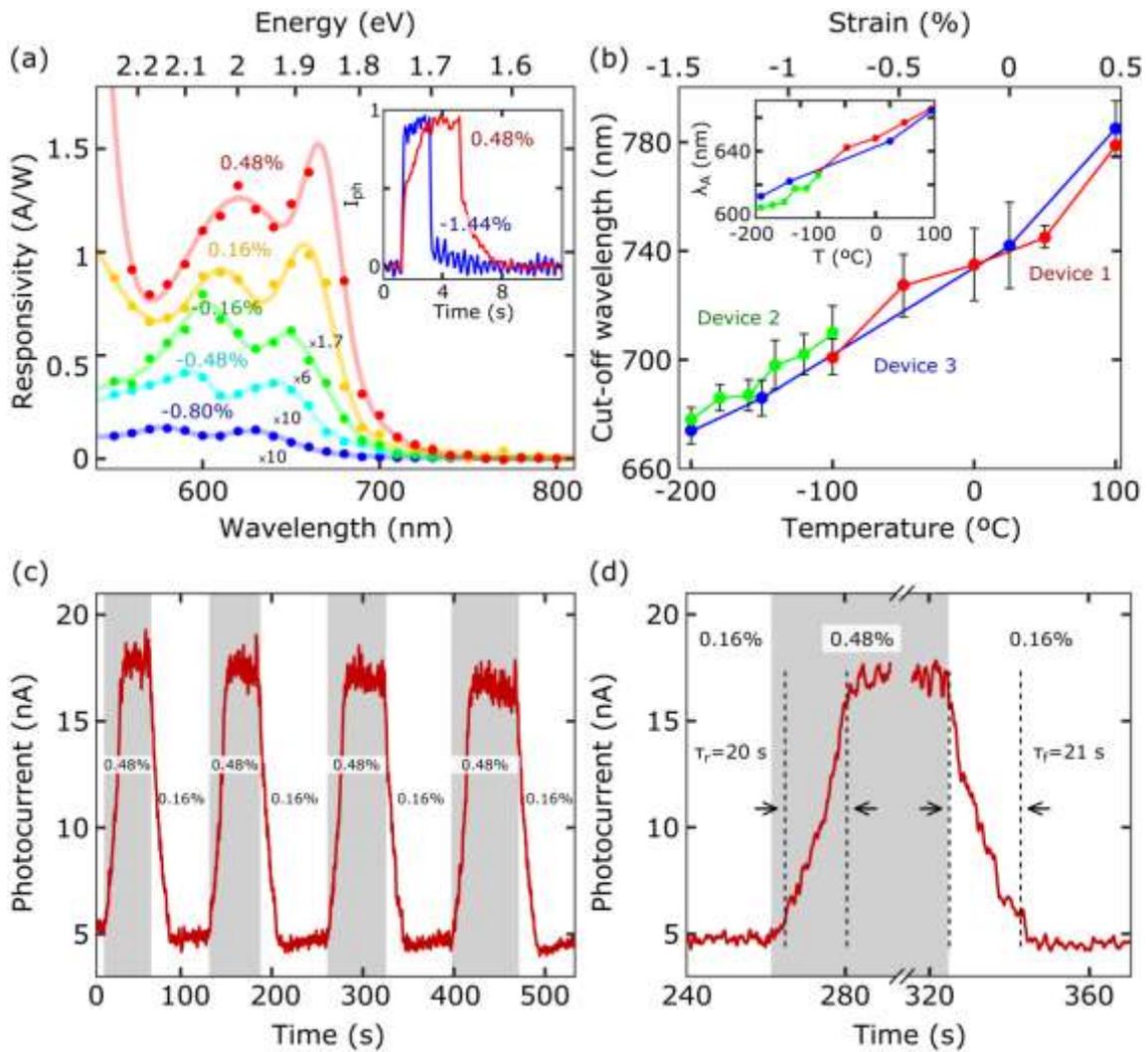

**Figure 2.** (a) Responsivity spectra of the single-layer MoS$_2$ photodetector #1 obtained by measuring under 5 different strains applied (from -0.80% to 0.48%). Each dot corresponds to the value measured under light power of 8 mW/cm$^2$ and applying a bias voltage of 10 V. Note that the responsivity values for +0.16%, -0.16%, -0.48% and -0.80% have been multiplied by 1.7, 6, 10 and 10 respectively to facilitate the comparison between the spectra. Inset: Response time for different strains applied. (b) Cut-off wavelengths extracted from the responsivity spectra of three single-layer MoS$_2$ photodetectors (#1, #2 and #3) at different strain ranges. Inset: Exciton A wavelengths extracted from the same spectra. (c) Response time to strain, being the OFF state 0.16% of strain applied and ON state 0.48% of strain applied, the cycles are measured with an applied voltage of 5 V and the 740 nm light source density power is 5 mW/cm$^2$. (d) Zoom in the third strain cycle of (c) in order to appreciate the rise and fall time of the device, estimated with the 10%-90% criterion.

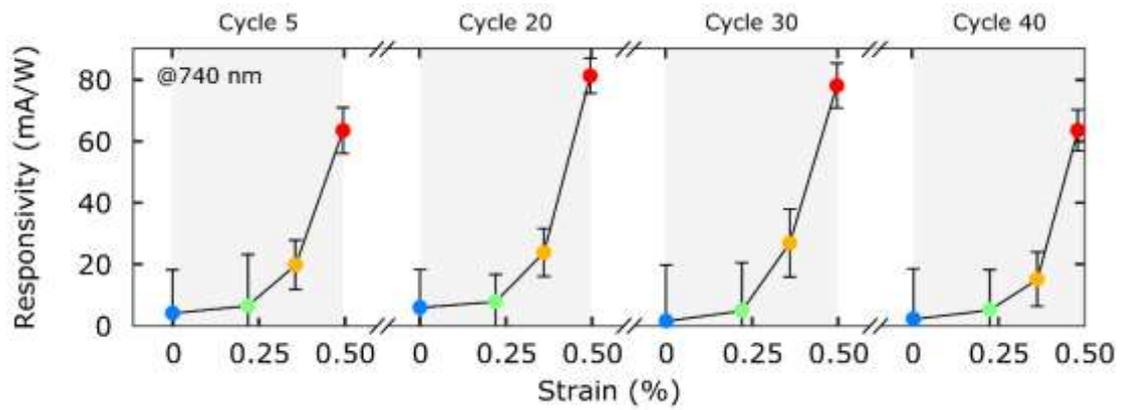

**Figure 3.** Photodetector #4 responsivities for 740 nm LED light with light power of 12 mW/cm$^2$ and applying a bias voltage of 10 V measured for several strain cycles.

Supplementary Material for:

# A strain tunable single-layer MoS$_2$ photodetector

*Patricia Gant[1], Peng Huang[1,2], David Pérez de Lara[3], Dan Guo[2], Riccardo Frisenda[1], Andres Castellanos-Gomez[1,*]*

[1]Materials Science Factory, Instituto de Ciencia de Materiales de Madrid (ICMM-CSIC), Madrid E-28049, Spain.

[2]State Key Laboratory of Tribology, Tsinghua University, Beijing 100084, China.

[3]Instituto Madrileño de Estudios Avanzados en Nanociencia (IMDEA-nanociencia),

Campus de Cantoblanco, Madrid, Spain.

Content of the Supplementary Material:

**Scanning micro-reflectance maps**

**Finite element analysis to determine the strain transfer**

**Strain calibration**

**Disentangling temperature change from biaxial strain**

**Scanning photocurrent measurements**

**Examples of strain tuning of the current vs. voltage characteristics and response time measurements**

**Characteristics of other devices**

**Strain tuning of the power-dependent photocurrent generation**

**Scanning micro-reflectance maps:**

Figure S1 shows micro-reflectance maps acquired on a $MoS_2$ device, fabricated onto polycarbonate (PC), at different temperatures. The maps show the spatial variation of the A exciton peak energy which gives an estimation of the strain spatial distribution in the device. The results show how the strain transfer from the thermal expansion of the substrate provides a very uniform strain distribution.

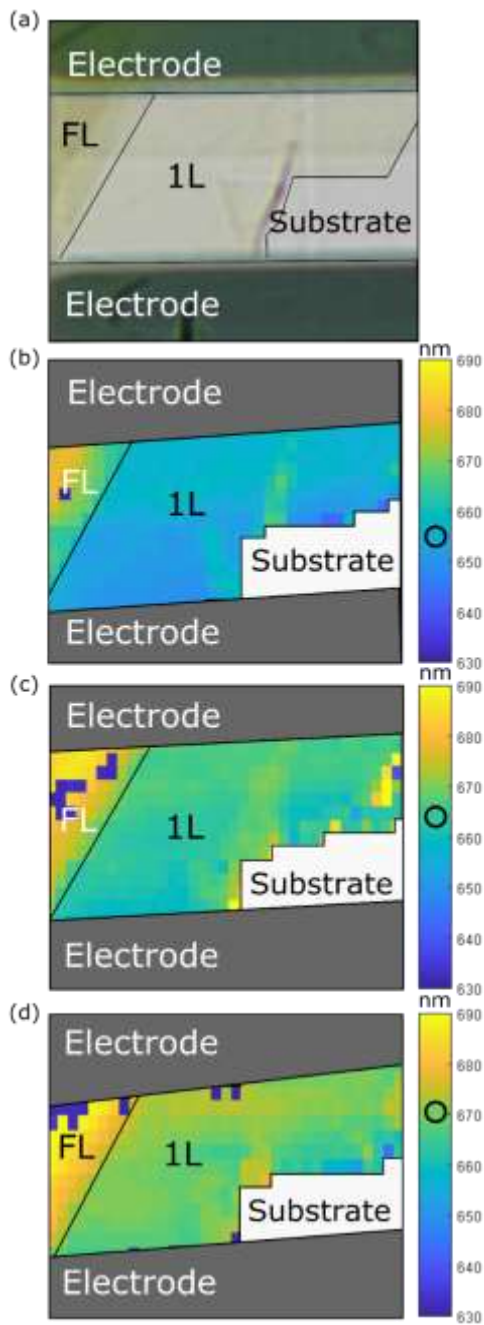

**Figure S1.** (a) Optical image of the sample to facilitate the identification of the different areas. (b-d) Exciton A energy map extracted from the differential reflectance excitonic resonance peak at 0% (b), 0.22% (c) and 0.35% (d) of strain.

**Finite element analysis to determine the strain transfer:**

To understand the role of the substrate in the transfer of strain to single-layer TMDCs deposited on top, we performed a three-dimensional axisymmetric finite element analysis (FEA).[39] The model consists of single-layer MoS$_2$ with a thickness of 0.7 nm and Young's modulus $E_{MoS2}$ = 246 GPa (Ref. [51] of the main text), placed on PDMS substrate with a thickness of 1000 µm. The FE model mesh was determined through a series of convergence studies. The interface between the MoS$_2$ flake and the substrate is modelled using perfect bonding. The calculations were performed using the commercial FE software COMSOL Multiphysics (version 5.2). In each step of the simulation we let the substrate expand thanks to thermal expansion and we extract the total expansion induced in the MoS$_2$ flake.

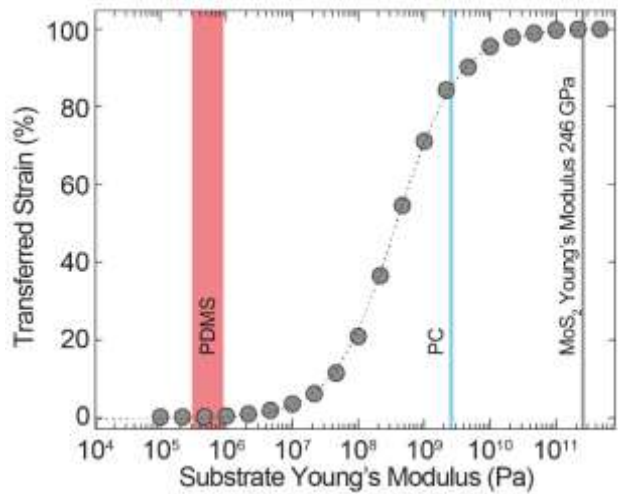

**Figure S2.** Finite element calculation of a biaxial strain test sample consisting of a 100-µm-thick substrate and single-layer MoS$_2$ (0.7 nm thickness) on the substrate. The curve represents the maximum transferred strain in MoS$_2$ as a function of substrate's Young's Modulus.

**Strain calibration:**

The technique used to calibrate the strain applied to the PC substrate with the temperature is explained in detail in Reference [39] of the main text. Briefly, we pattern (with photolithography) a periodic array of pillars on the substrate that are used as markers to facilitate the determination of the expansion/contraction of the substrate through optical inspection. Optical microscopy images are acquired at different temperatures and the relative distance between the pillars are determined through an image recognition script that identifies the coordinates of the pillars (Figure S3).

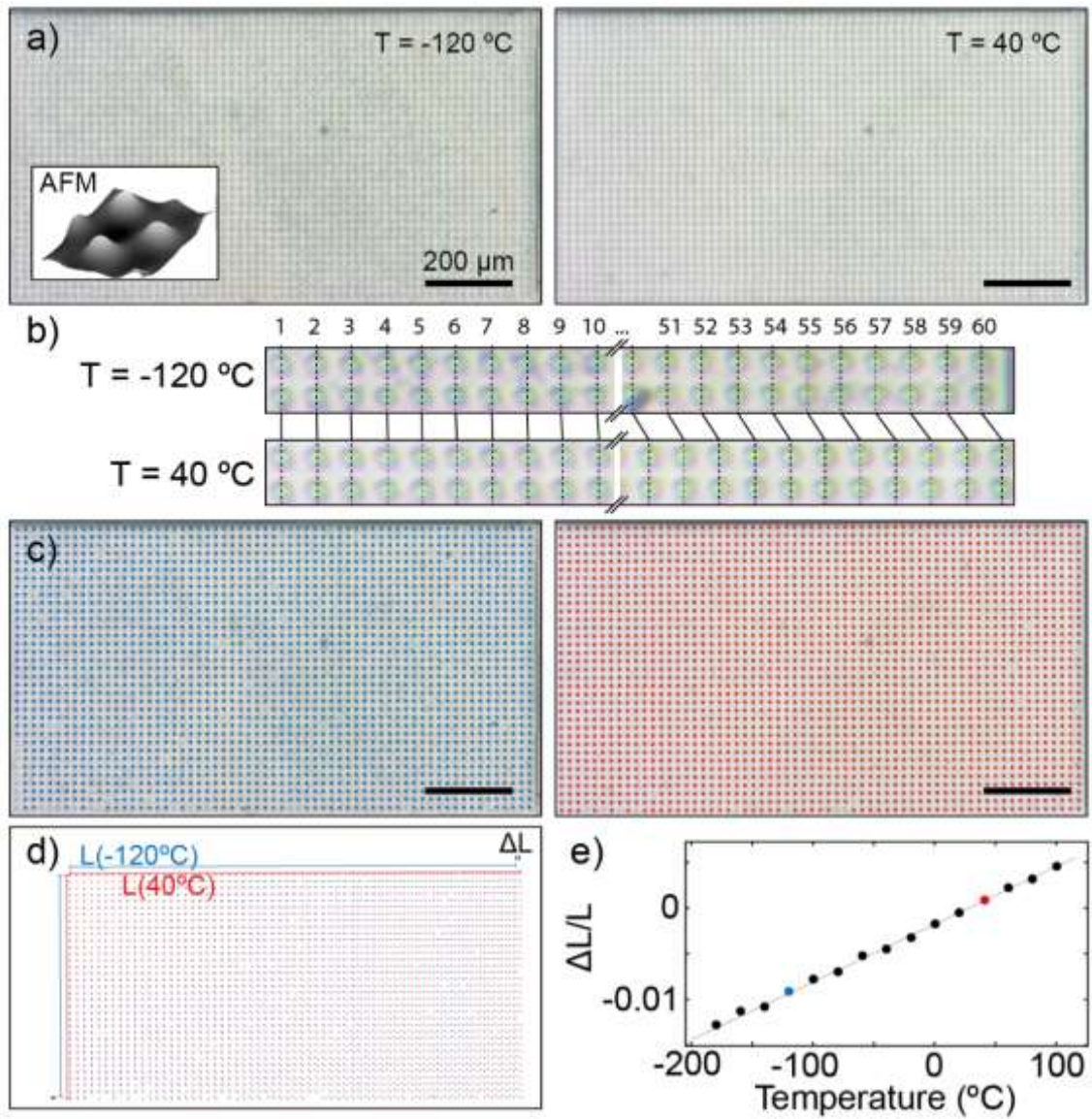

**Figure S3.** (a) Comparison between two microscopy images of a polycarbonate (PC) substrate, with lithographic photoresist pillars on top, acquired at -120ºC and +40ºC respectively. (b) Compares the position of the pillars at the two temperatures where a marked compression on the -120ºC is visible. (c) Shows an example where automatic recognition of the position of the pillars is demonstrated to facilitate even further the direct comparison of the pictures acquired at different temperatures (and the direct determination of the thermal expansion/compression of the substrate, panel (d). (e) Temperature dependence of the biaxial expansion/compression of the PC substrate. The solid line is the linear fit of the experimental data (the slope is $6.4 \cdot 10^{-3}$ %/ºC).

**Disentangling temperature change from biaxial strain:**

We have compared the performance of devices fabricated on polycarbonate (PC) which has a large thermal expansion with devices fabricated onto $SiO_2/Si$ that has a negligible thermal expansion coefficient. In this way we can effectively distinguish if an observation is due to the biaxial strain or is due the change of temperature.

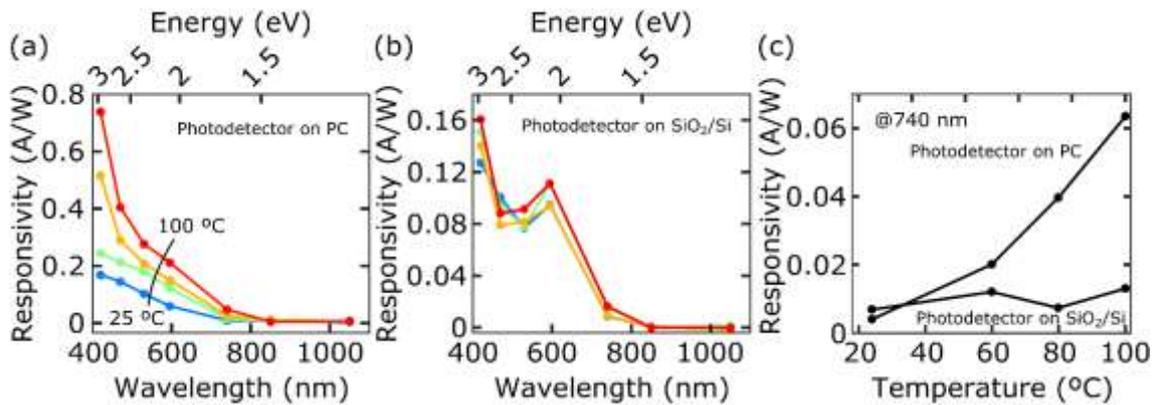

**Figure S4.** (a) Responsivity spectra of the single-layer MoS$_2$ photodetector #4 on PC obtained at 25ºC (blue), 60ºC (green), 80ºC (orange) and 100ºC (red). These values correspond to no strain (blue), 0.22% of strain (green), 0.35% (orange) and 0.48% (red). Each dot corresponds to the value measured under light power of 12 mW/cm$^2$ and applying bias voltage of 10 V. (b) Responsivity spectra of the single-layer MoS$_2$ photodetector on SiO$_2$/Si obtained at 25ºC (blue), 60ºC (green), 80ºC (orange) and 100ºC (red). Each dot corresponds to the value measured under light power of 12 mW/cm$^2$ and applying bias voltage of 10 V. (c) Responsivity measured with a 740 nm LED light at the same conditions as (a) and (b) for photodetector #4 on PC (application of strain with the temperature) and photodetector on SiO$_2$ (no strain application with the temperature).

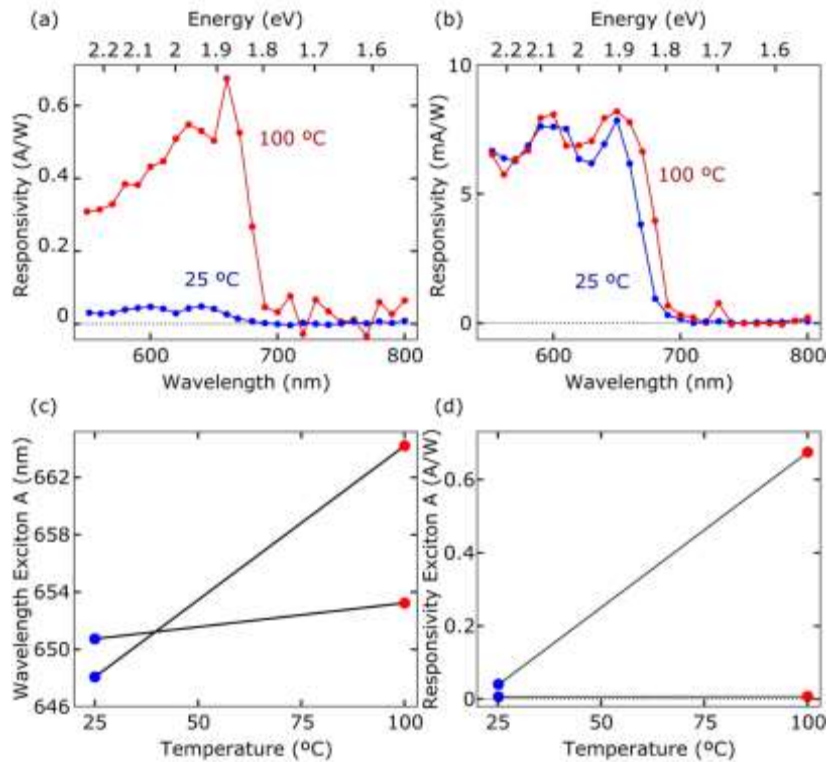

**Figure S5.** (a) Responsivity spectra of the single-layer MoS$_2$ photodetector #3 on PC obtained at 25ºC (blue) and 100ºC (red). Each dot corresponds to the value measured under illumination with monochromatic light with a power of 8 mW/cm$^2$ and applying bias voltage of 10 V. (b) Responsivity spectra of the single-layer MoS$_2$ photodetector on SiO$_2$/Si obtained at 25ºC (blue) and 100ºC (red). Each dot corresponds to the value measured under illumination with monochromatic light with a power of 8 mW/cm$^2$ and applying bias voltage of 10 V. (c) Wavelength of the exciton A for 25 ºC and 100 ºC, extracted from a gaussian fit. (d) Responsivity in the exciton A for 25 ºC and 100 ºC, extracted from a gaussian fit.

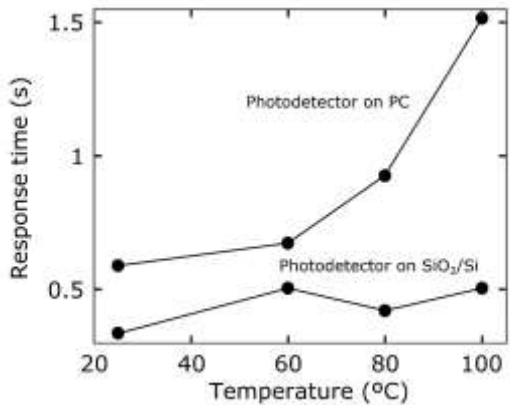

**Figure S6.** Fall response time of the photodetector on $SiO_2/Si$ and photodetector #4 on PC extracted for four different temperatures, measured with 420 nm LED light at bias voltage of 10 V.

**Scanning photocurrent measurements:**

We employed scanning photocurrent measurements at different temperatures to determine the strain dependence of the Schottky barrier height. Figure S7 displays the photocurrent maps obtained by scanning a diffraction limited spot (of 650 nm of wavelength) over the sample. Figure S8 shows horizontal linecuts where the photocurrent generation at the metal-semiconductor interface can be easily resolved. By repeating the measurements at different bias voltages one can estimate the barrier height through the voltage needed to null the photocurrent at the metal-semiconductor interface.

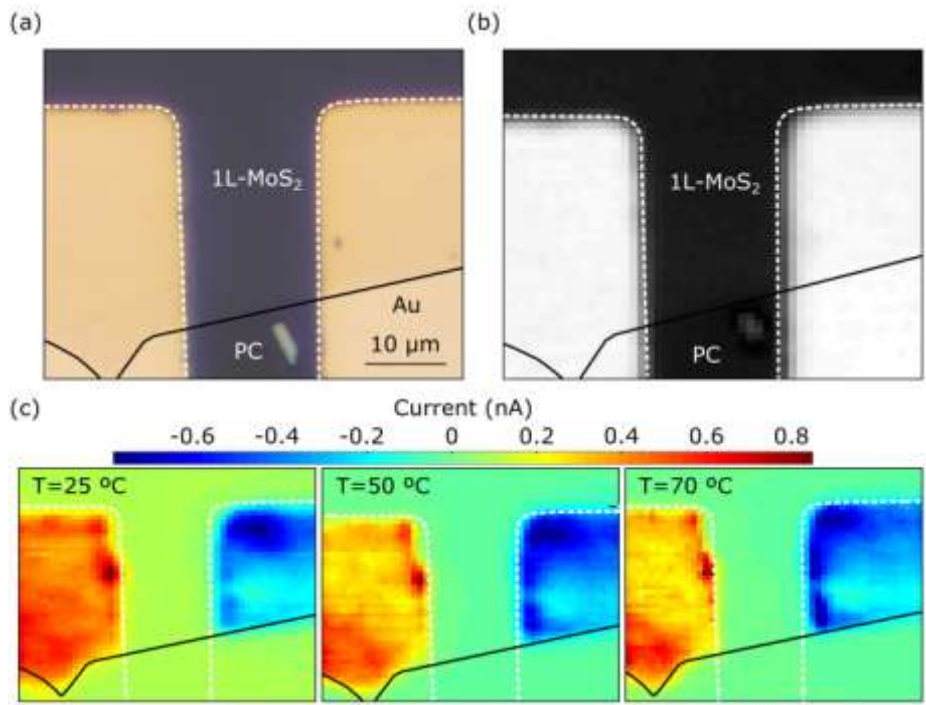

**Figure S7.** (a) Optical microscopy image in reflection mode of the single-layer $MoS_2$ photodetector #7 on polycarbonate (PC). (b) Reflection map (reflected intensity of the laser) of the same photodetector acquired during the scanning photocurrent measurement using a laser of 650 nm focalized on a diffraction limited spot. (c) Current maps of the same photodetector at different temperatures. The lines are placed using the reflection maps to indicate the position of the flake and the electrodes.

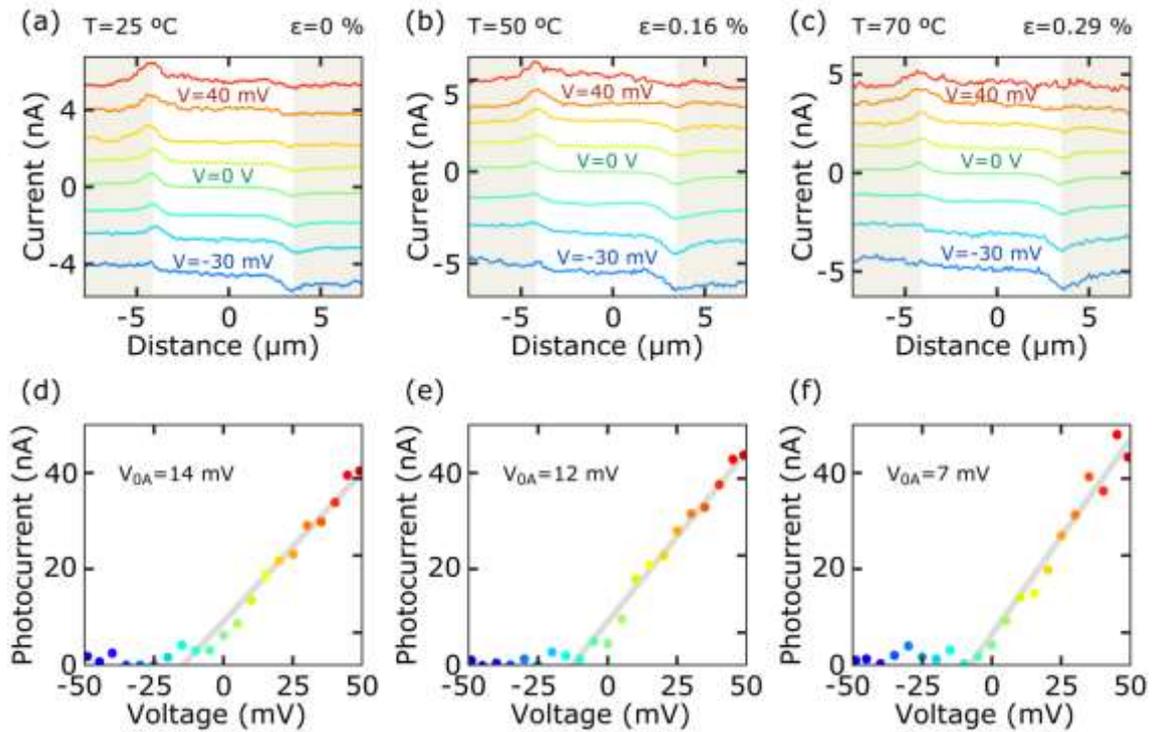

**Figure S8.** (a), (b), (c) Current linecuts in the single-layer MoS$_2$ photodetector #7 on PC measured for different bias voltages applied (from -50 mV to 50 mV, only few selected bias voltages are displayed for clarity) in the nearby area of the electrodes at different temperatures. (d), (e), (f) Photocurrent measured in the left electrode at different bias voltages. The grey line is the linear fit of photocurrent datapoints with values above the noise level. The cross of this line with the horizontal axis represent an estimation of Schottky barrier height at each temperature.

**Examples of strain tuning of the current vs. voltage characteristics and response time measurements:**

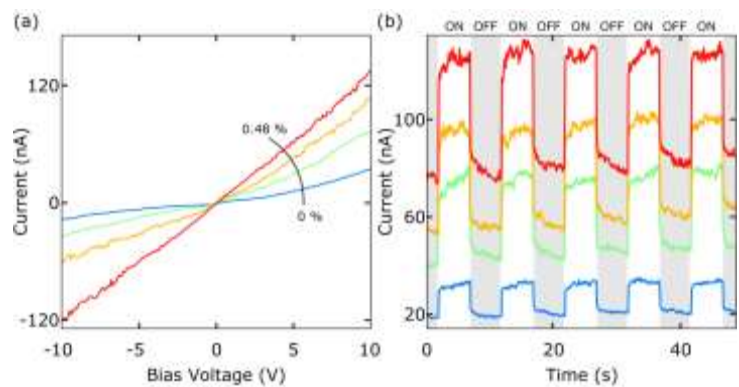

**Figure S9.** (a) Single-layer MoS$_2$ photodetector #4 current vs voltage characteristics measured under different strain applied (from 0 % (blue) to 0.48 % (red)) by illuminating 420 nm LED light with light power of 12 mW/cm$^2$. (b) Response time measured with an applied voltage of 10 V and the 420 nm light source density power is 12 mW/ cm$^2$. Each curve corresponds to different strain value (0 % (blue) to 0.48 % (red)).

**Characteristics of other devices:**

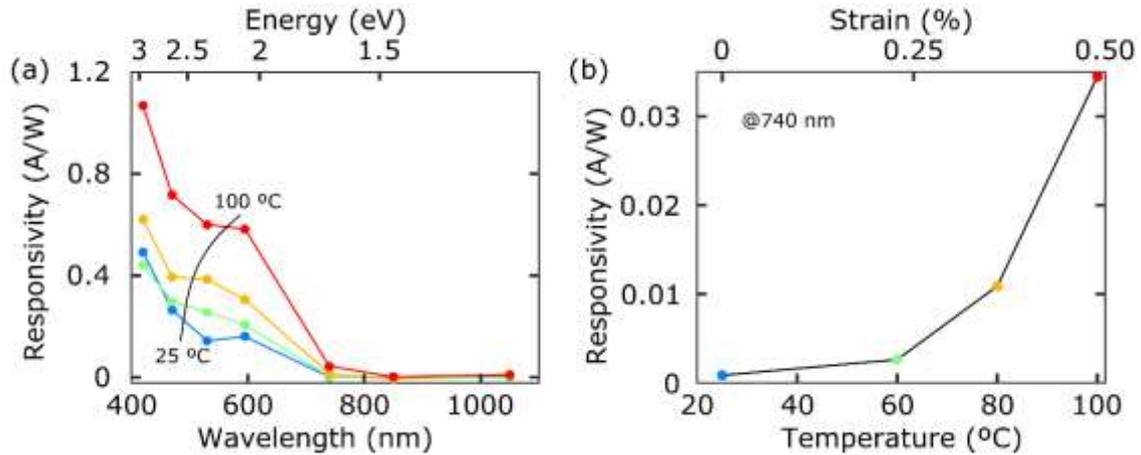

**Figure S10.** (a) Responsivity spectra of the single-layer MoS$_2$ photodetector #5 obtained for no strain (blue), 0.22% of strain (green), 0.35% (orange) and 0.48% (red). Each dot corresponds to the value measured under light power of 12 mW/cm$^2$ and applying a bias voltage of 10 V. (b) Responsivity measured with a 740 nm LED light at the same conditions as (c).

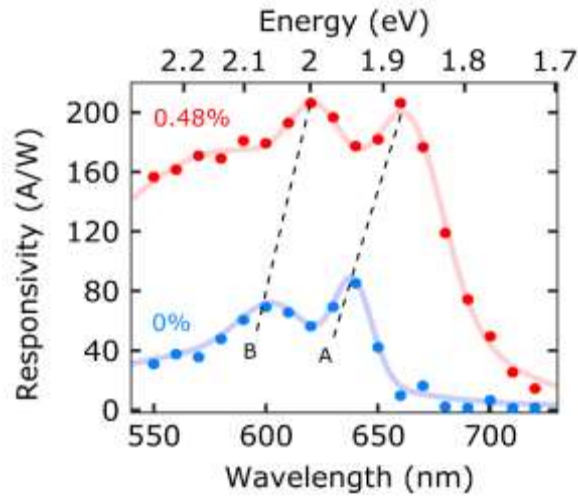

**Figure S11.** Responsivity spectra of the single-layer MoS$_2$ photodetector #6 obtained without strain (blue) and with 0.48% of strain applied (red). Each dot corresponds to the value measured under a light power of 80 μW/cm$^2$ and applying a bias voltage of 10 V.

**Strain tuning of the power-dependent photocurrent generation:**

From the power dependence of the photocurrent we can extract the power-law exponent α that provides information about the photocurrent generation mechanism. For tensile strain the exponent is well-below 1 indicating a strong photogating effect. For compressive strain, although the exponent is still below 1 it increased sizeably with respect to the case of the tensioned state. This is in agreement with the reduced responsivity and fast response of the compressed devices that points to photoconductive mechanism as taking over.

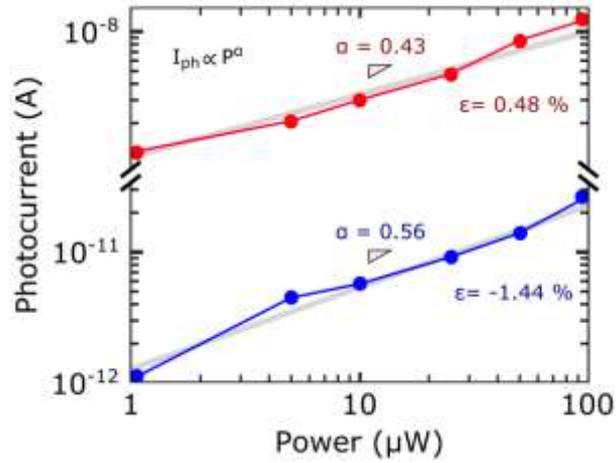

**Figure S12.** Photocurrent *vs.* power for different applied strain levels on the single-layer MoS$_2$ photodetector #3. The photocurrent is extracted from a response time measurement under a bias voltage of 10 V and illuminating with a LED source with a wavelength of 505 nm. The grey solid lines correspond to the best fit to determine the power-law exponent.